\documentclass{ws-procs10x7}
\usepackage{balance}

\newcolumntype{d}[1]{D{.}{.}{#1}}

\def\Journal#1#2#3#4{{\it #1} {\bf #2}, #3 (#4)}

\makeindex
\def\be{\begin{eqnarray}}
\def\ee{\end{eqnarray}}
\newcommand{\beqar}{\begin{eqnarray*}}
\newcommand{\eeqar}{\end{eqnarray*}}
\def\ba{\begin{array}}
\def\ea{\end{array}}
\def\p{\phi}
\def\vp{\varphi}

\def\pa{\partial}

\def\G{{\cal G}}
\def\B{{\cal B}}
\def\A{{\cal A}}
\def\X{{\cal X}}

\def\D{^{(5)}}
\begin{document}

\title{BLACK RING SOLUTIONS IN 5D HETEROTIC STRING THEORY:\\
A FULL FIELD CONFIGURATION}

\author{A. HERRERA--AGUILAR$^*$ \and H. R. MARQUEZ--FALCON$^\ddag$}

\address{Instituto de F\'\i sica y Matem\'aticas, Universidad Michoacana de San Nicol\'as de
Hidalgo, \\
Edificio C--3, Ciudad Universitaria, CP 58040 Morelia,
Michoac\'an, M\'exico\\$^*$E-mail: herrera@ifm.umich.mx \
$^\ddag$E-mail: hugo@ifm.umich.mx}




\twocolumn[\maketitle\abstract{We show how to obtain a
non--supersymmetric black ring configuration in the framework of
5D Heterotic String Theory using the matrix Ernst potential (MEP)
formalism, which enables us to include non--trivial dilaton, gauge
and antisymmteric tensor fields.}  ]

\section{3D effective action and matrix Ernst potentials}

We shall begin with the 5D effective action of the heterotic
string at tree level \be S\D=\int\!d^5x\mid G\D
\mid^{\frac{1}{2}}e^{-\phi\D}\left(R\D+
\right. \nonumber\\
\phi\D_{;M}\,\phi^{(5);M}\!-\!\frac{1}{12}\,H\D_{MNP}\,H^{(5)MNP}-\\
\left.\frac{1}{4}\,F^{(5)I}_{MN}\,F^{(5)IMN}\right), \nonumber\ee
where $G\D_{MN}$ is the 5D metric, $\phi\D$ is the 5D dilaton,
$F^{(5)I}_{MN}=\pa_MA^{(5)I}_N-\pa_NA^{(5)I}_M$ and
$H\D_{MNP}\!\!=\!\!\pa_MB\D_{NP}\!\!-\!\frac{1}{2}A^{(5)I}_MF^{(5)I}_{NP}\!+
\mbox{\rm cycl. perms. of M,N,P.}$ Here $B\D_{MN}$ is the
anti--symmetric Kalb--Ramond field and $A^{(5)I}_M$
($I=1,\,2,\,...,n$) is a set of $U(1)$ vector fields. In Ref.
\cite{ms} it was shown that after imposing two commuting Killing
vectors, the resulting stationary theory possesses the
$U$--duality symmetry group $SO(3,3+n)$ and describes 3D gravity
\begin{eqnarray}
g_{\mu\nu}\!\!=\!e^{-2\p}\!\left(G\D_{\mu\nu}\!-
\!G\D_{m+3,\mu}G\D_{n+3,\nu}G^{mn}\right),
\end{eqnarray}
coupled to the following set of three--fields:
\noindent a) scalar
fields
\be
\ba{rcl} G\!&\equiv&\!G_{mn}\!= \!G\D_{m+3,n+3},\cr
B\!&\equiv&\!B_{mn}\!= \!B\D_{m+3,n+3}, \cr A\!&\equiv&\!A^I_m\!=
\!A^{(5)I}_{m+3},\cr \phi&=&\phi\D\!-\!\frac{1}{2}{\rm
ln|det}\,G|, \ea
\ee
\noindent b)tensor field (we set it to zero in this report)
\begin{eqnarray}
B_{\mu\nu}\!\!&=&\!B\D_{\mu\nu}\!\!-\!4B_{mn}A^m_{\mu}A^n_{\nu}\!\\
&-&\!2\!\left(A^m_{\mu}A^{2+m}_{\nu}\!-\!A^m_{\nu}A^{2+m}_{\mu}\right)\equiv
0 \nonumber, \end{eqnarray}
\noindent c)vectors
$A^{(a)}_{\mu}\!\!=\!
\left(\!(A_1)^m_{\mu},\!(A_2)^{2+m}_{\mu},\!(A_3)^{4+I}_{\mu}\!\right)$
\begin{eqnarray}
(A_1)^m_{\mu}\!&=&\!\frac{1}{2}G^{mn}G\D_{n+3,\mu},\nonumber \\
(A_3)^{4+I}_{\mu}\!&=&\!-\frac{1}{2}A^{(5)I}_{\mu}\!+\!A^I_nA^n_{\mu},
\\
(A_2)^{2+m}_{\mu}\!&=&\!\frac{1}{2}B\D_{m+3,\mu}\!\!-
\!B_{mn}A^n_{\mu}\!+\!\frac{1}{2}A^I_{m}A^{4+I}_{\mu} \nonumber
\end{eqnarray}
where the subscripts $m,n\!=\!1,2$; and $a\!=\!1,...,4+n$. All
vector fields in three dimensions can be dualized on--shell as
follows:
\begin{eqnarray}
\ba{l}
\nabla\!\times\!\overrightarrow{A_1}\!=\frac{1}{2}e^{2\p}G^{-1}\times\cr
\left[\nabla u\!+\!(B\!+\!\frac{1}{2}AA^T)\nabla v\!+ \!A\nabla
s\right], \cr
\nabla\!\times\!\overrightarrow{A_3}\!=\frac{1}{2}e^{2\p} (\nabla
s\!+\!A^T\nabla v)\!+\!A^T\nabla\!\times\!\overrightarrow{A_1},
\cr
\nabla\!\times\!\overrightarrow{A_2}\!=\frac{1}{2}e^{2\p}G\nabla
v- \cr
(B\!+\!\frac{1}{2}AA^T)\nabla\!\times\!\overrightarrow{A_1}+A\nabla\!\times\!\overrightarrow{A_3}.
\ea \label{dual}
\end{eqnarray}
Thus, the effective stationary theory describes gravity
$g_{\mu\nu}$ coupled to the scalars $G$, $B$, $A$, $\p$ and
pseudoscalars $u$, $v$, $s$. These matter fields can be arranged
in the following pair of matrix Ernst potentials \cite{hkMEP}: \be
\X\!=\!\left(\ba{ccc}
\!-\!e^{\!-2\p}\!+\!v^T\!Xv\!+\!v^T\!As\!+\!\frac{1}{2}s^T\!s&&v^T\!X\!-\!u^T
\cr Xv+u+As&&X \ea \right), \nonumber \ee \be \A=\left( \ba{c}
s^T+v^TA \cr A \ea \right), \ee where $X=G+B+\frac{1}{2}AA^T$.
These matrices have dimensions $3\times3$ and $3\times n$,
respectively. Thus, the effective stationary theory adopts the
following form in terms of the MEP:
\begin{eqnarray}
\!S^3\!=\!\int\!d^3x\mid\!g\!\mid^{\frac{1}{2}}\!\left\{\!-R\!+\!
{\rm Tr}\!\left[\!\frac{1}{2}\nabla\A^T\G^{-1}\nabla\A\right.\right.\!\!+\!\nonumber \\
\left.\left.\!\frac{1}{4}\left(\nabla \X\!-\!\nabla
\A\A^T\right)\!\G^{-1}\!\left(\nabla\X^T\!-\!\A\nabla\A^T\right)\!\G^{-1}\right]\right\},
\nonumber
\end{eqnarray} with $\X=\G+\B+\frac{1}{2}\A\A^T$, \, and hence,
$\G=\frac{1}{2}\left(\X+\X^T-\A\A^T\right)$ and
$\B=\frac{1}{2}\left(\X-\X^T\right)$.

In \cite{hkMEP} it was also shown that there exist a map between
the stationary actions of the heterotic string and
Einstein--Maxwell theories: \be \X\longleftrightarrow -E, \qquad
\qquad \A\longleftrightarrow F, \ee where ${\it matrix\,\,
transposition}$ is also interchanged with ${\it complex\,\,
conjugation}$; $E$ and $F$ are the gravitational and
electromagnetic complex Ernst potentials of the stationary
Einstein--Maxwell theory \cite{e}.

In the language of the MEP the latter stationary action possesses
a set of isometries which has been clasified according to their
charging properties in Ref. \cite{hk5}. Among them, the so--called
normalized Harrison transformation (NHT) acts in a non--trivial
way on a seed spacetime: it allows us to construct charged string
vacua from neutral ones preserving the asymptotical values of the
initial fields. Namely, the matrix transformation
\begin{eqnarray}
\ba{l}
\A\!\rightarrow\!\left(\!1\!+\!\frac{1}{2}\Sigma\lambda\lambda^T\!\right)\!
\left(\!1\!-\!\A_0\lambda^T\!+\!\frac{1}{2}\X_0\lambda\lambda^T\right)^{-1}\!\times\cr
\left(A_0-\X_0\lambda\right)+\Sigma\lambda,\cr
\X\!\rightarrow\!\left(\!1\!+\!\frac{1}{2}\Sigma\lambda\lambda^T\!\right)\!
\left(\!1\!-\!\A_0\lambda^T\!+\!\frac{1}{2}\X_0\lambda\lambda^T\right)^{-1}\!\times\cr
\left[\X_0+\left(\A_0-\frac{1}{2}\X_0\lambda\right)\lambda^T\Sigma\right]
+\frac{1}{2}\Sigma\lambda\lambda^T\Sigma, \label{nht}\ea
\end{eqnarray}
where $\Sigma=diag(-1,-1;1,1,...,1)$ and $\lambda$ is an arbitrary
constant $3\times n$--matrix parameter, generates charged string
solutions (with non--zero potential $\A$) from the neutral seed
ones $\X_0\ne 0$, and $\A_0=0$. This solution--generating
technique allows us to generate the $U(1)^n$ electromagnetic
spectrum of the effective heterotic string theory starting with
just the bosonic spectrum of string theory.

\section{Charging a neutral black ring}

We can apply the NHT on a seed solution corresponding to the 5D
{\it neutral black ring} of Emparan and Reall \cite{er} (see also
Ref. \cite{ee}): \be \ba{l}\label{ring0}
ds^2=-\frac{F(x)}{F(y)}\left(dt+R\sqrt{\lambda\nu}(1+y)
d\psi\right)^2+\cr
\frac{R^2}{(x-y)^2}\left[-F(x)\left(G(y)d\psi^2+\frac{F(y)}{G(y)}
dy^2\right)+\right.\cr
\left.F(y)^2\left(\frac{dx^2}{G(x)}+\frac{G(x)}{F(x)}d\phi^2\right)\right]
\ea\ee with \be
  F(\xi) = 1 - \lambda\xi \, ,
\quad  G(\xi) = (1 - \xi^2)(1-\nu \xi).\nonumber \label{fandg}\ee
Here $R$ is the radius of the ring, $F(\xi)$ has the root
$\xi_1=\lambda^{-1}$, and $G(\xi)$ possesses three roots:
$\xi_2=-1\,,\quad \xi_3=+1\,,\quad \xi_4=\frac{1}{\nu}\,.$ Thus,
the solution has two dimensionless parameters $\lambda$ and $\nu$.

\noindent The variables $x$ and $y$ take values in \be -1\leq
x\leq 1\,,\ -\infty<y\leq-1\,,\ \lambda^{-1}<y<\infty\,.
\label{xyrange}\nonumber\ee

\noindent In order to balance forces in the ring, one must
identify $\psi$ and $\phi$ with equal period \be
\Delta\phi\!=\!\Delta\psi\!=\!\frac{4\pi
\sqrt{F(-1)}}{|G'(-1)|}\!=\!\frac{2\pi\sqrt{1+\lambda}}{1+\nu}\,.
\label{phsiperiod} \ee

\noindent This eliminates the conical singularities at the
fixed-point sets $y=-1$ and $x=-1$ of the Killing vectors
$\partial_\psi$ and $\partial_\phi$, respectively.

For avoiding conical singularities at $x=+1$, we have two cases:

\noindent 1) By fixing \be \lambda=\lambda_c\equiv
\frac{2\nu}{1+\nu^2}\qquad {\rm (black\; ring)} \label{lambring}
\ee makes the circular orbits of $\partial_\phi$ close off
smoothly also at $x=+1$. Then $(x,\phi)$ parameterize a
two--sphere, $\psi$ parameterizes a circle, and the solution
describes a {\bf black ring}.

\noindent 2) If we set \be \lambda=1\qquad {\rm (black\; hole)}
\label{lambhole}\ee the orbits of $\partial_\phi$ do not close at
$x=+1$. Then $(x,\phi,\psi)$ parameterize an $S^3$ at constant
$t,y$. The solution describes the {\bf black hole} of Myers and
Perry with a single rotation parameter \cite{mp}.

\noindent Both for black holes and black rings, $|y|=\infty$ is an
ergosurface, $y=1/\nu$ is the event horizon, and as
$y\to\lambda^{-1}$ an inner, spacelike singularity is reached from
above.

The parameter $\nu$ varies in the range \be\label{nurange}
0\leq\nu<1\,. \ee

As $\nu\to 0$ we recover a non--rotating black hole, or a very
thin black ring. At the opposite limit, $\nu\to 1$, both the black
hole and the black ring result into the same solution with a naked
ring singularity. If one considers $\lambda$ as an independent
parameter which can be eventually fixed to the equilibrium value
$\lambda_c$. If $\lambda$ adopts values different from
(\ref{lambring}) and (\ref{lambhole}), then whenever
$\nu<\lambda<1$ one finds a black ring solution regular on and
outside the horizon, except for a conical singularity on the disk
bounded by the inner rim of the ring ($x=+1$). If
$\nu<\lambda<\lambda_c$ then the ring is rotating faster than the
equilibrium value, and there is a conical deficit balancing the
excess centrifugal force. If instead $\lambda_c<\lambda<1$ then
the rotation is too slow and a conical excess appears. If
$\lambda\leq\nu$ the horizon is replaced by a naked singularity.
Finally, the solution with $\lambda=\lambda_c$ is the equilibrium,
or balanced black ring.

The mass, spin, area, temperature and the angular velocity at the
horizon for the black ring are given by \be M_0=\frac{3\pi
R^2}{4G}\frac{\lambda(\lambda+1)}{\nu+1}, \label{mandj}\ee \be
J_0=\frac{\pi
R^3}{2G}\frac{\sqrt{\lambda\nu}(\lambda+1)^{5/2}}{(1+\nu)^2}\,,
\label{mandj}\ee \be \A_0 =
  8 \pi^2 R^3 \frac{\lambda^{1/2} (1+\lambda)(\lambda - \nu)^{3/2}}
       {(1+\nu)^2 (1-\nu)}\,,
\label{areatemp}\ee \be
  T_0 = \frac{1}{4\pi R}\frac{1-\nu}
  {\lambda^{1/2} (\lambda - \nu)^{1/2} }
     \, ,
\label{areatemp}\ee

\be \Omega_0=\frac{1}{R}\sqrt{\frac{\nu}{\lambda(1+\lambda)}}\,.
\label{omega0}\ee

Looking at the dimensionless quantity \be \frac{27\pi}{32
G}\frac{J_0^2}{M_0^3}=\left\{
\begin{array}{ll}
 \displaystyle{\frac{2\nu}{\nu+1}}\ \textrm{(black hole)}\\
& \\
\displaystyle{\frac{(1+\nu)^3}{8\nu}}\ \textrm{(balanced black
ring)}
\end{array} \right.
\label{jonm}\ee one sees that for black holes it grows
monotonically from $0$ to $1$, while for (equilibrium) black rings
it is infinite at $\nu=0$, decreases to a minimum value $27/32$ at
$\nu=1/2$, and then grows to $1$ at $\nu=1$. This implies that in
the range \be \frac{27}{32}\leq\frac{27\pi}{32
G}\frac{J_0^2}{M_0^3}<1 \label{nonuniquej}\ee there exist {\bf one
black hole} and {\bf two black rings} with the same value of the
spin for fixed mass.

This regime of {\bf non--uniqueness} occurs when the parameter
$\nu$ takes values in \be \sqrt{5}-2\leq \nu <1
\label{nonuniquenu} \ee for equilibrium black rings, and in \be
\frac{27}{37}\leq \nu<1 \label{nonuniquenubh} \ee for black holes.

\section{Solution--generating Technique}
This 5D {\it neutral black ring} constitutes a solution of the
heterotic string theory when \be \p\D=0, \quad A^{(5)I}_M=0, \quad
B\D_{MN}=0, \ee and the corresponding MEP read \be \X=\left(
\ba{cc} -e^{-2\p}&-u^T \cr u&G \ea \right), \quad \A=0. \ee After
applying the NHT one obtains a new field configuration that
corresponds to a {\bf charged black ring} with non--trivial
dilaton, Kalb--Ramond and electromagnetic fields. Similar results
have been applied to 5D and 4D theories in Refs.
\cite{aha0}--\cite{bha}. These fields can be extracted from the
respective components of the generating matrix Ernst potentials
\be \X= \left( \ba{cc} \X_{11}&\X_{12} \cr \X_{21}&\X_{22} \ea
\right), \quad \A= \left( \ba{c} \A_{1n} \cr \A_{2n} \ea \right).
\ee

Thus, the 5D metric adopts the form \be\ba{l}
ds^2=G_{MN}dx^{M}dx^{N}=\cr G_{mn}\left(dx^{m+3}\!+\!\omega^{(m)}
d\vp\right)\left(dx^{n+3}\!+\!\omega^{(n)} d\vp\right)\!+\!\cr
e^{2\p}g_{\mu\nu}dx^{\mu}dx^{\nu}, \ea\nonumber\ee where the
metric components $G\equiv G_{mn}$ and $e^{2\p}$ read
$$G=\frac{1}{2}\left(\X_{22}+\X_{22}^T-\A_{2n}\A_{2n}^T\right)$$
and \be\ba{l}e^{-2\p}=-\X_{11}+\frac{1}{2}\A_{1n}\A_{1n}^T+\cr
\frac{1}{8}
\left(\X_{12}+\X_{21}^T-\A_{1n}\A_{2n}^T\right)G^{-1}\times\cr
\left(2G-\A_{2n}\A_{2n}^T\right)G^{-1}\left(\X_{21}+\X_{12}^T-\A_{2n}\A_{1n}^T\right),
\nonumber\ea\ee and $g_{\mu\nu}$ is the spatial 3D metric.

The remaining 5D fields can be obtained from the matrix
expressions
$$B=\frac{1}{2}\left(\X_{22}-\X_{22}^T\right),$$
$$A=\A_{2n},$$
$$v=\frac{1}{2}G^{-1}\left(\X_{21}+\X_{12}^T-\A_{2n}\A_{1n}^T\right),$$
$$u=\frac{1}{2}\X_{22}^TG^{-1}\left(\X_{21}+\X_{12}^T-\A_{2n}\A_{1n}^T\right)-\X_{12}^T,$$
$$s=\A_{1n}^T-\frac{1}{2}\A_{2n}^TG^{-1}\left(\X_{21}+\X_{12}^T-\A_{2n}\A_{1n}^T\right),$$
remembering that in order to recover the non--trivial components
of the 3D vector fields one must make use of the dualization
relations (\ref{dual}) for the pseudoscalar fields.

\section*{Acknowledgments}
Both authors thank the organizers of the ICHEP06 Conference for
providing a warm atmosphere plenty of interesting and useful
discussions in Moscow. This research was supported by grants
CIC-UMSNH-4.16 and CONACYT-F42064.

\end{document}